\documentclass[
    ,draft            
  ]
  {aipproc}

\layoutstyle{8x11single}

\def\wcen{$\omega$ Cen}

\def\rc{r$_c$}

\def\eg{{e.g.}}

\def\ie{{i.e.}}

\def\x{$\times$}
\def\about{$\sim$}
\def\simlt{$\buildrel{<}\over \sim$}

\def\asec{$''$}
\def\amin{$'$}
\def\secspt{$\buildrel{\prime\prime}\over .$}

\def\msun{${\cal M}_{\odot}$}

\def\yr-1{yr$^{-1}$}

\def\bwfpc2{$B_{439}$}
\def\b{$B_{435}$}

\def\r{$R_{625}$}
\def\i{$I_{814}$}

\def\br{$B_{435}-R_{625}$}

\def\hrwfpc2{$H\alpha-R_{675}$}
\def\hr{$H\alpha-R_{625}$}

\def\ha{H$\alpha$}

\def\w{$\omega$}

\def\Chandra{$\it Chandra$}
\def\HST{$\it HST$}

\def\apj{\rm ApJ}
\def\apjl{\rm ApJL}

\def\aj{\rm AJ}
\def\aap{\rm A\&A}
\def\mnras{\rm MNRAS}
\def\pasp{\rm PASP}
\def\araa{\rm ARA\&A}
\def\aaps{\rm A\&AS}

\begin{document}

\title{A Deep Multiwavelength View of Binaries in $\omega$ Centauri}



\author{Daryl Haggard}{
  address={Center for Interdisciplinary Exploration and Research in Astrophysics, Northwestern University, 2145 Sheridan Road, Evanston, IL 60208, USA; dhaggard@northwestern.edu}
}

\author{Adrienne M. Cool}{
  address={Department of Physics and Astronomy, San Francisco State University, 1600 Holloway Avenue, San Francisco, CA 94132, USA}
}

\author{Tersi Arias}{
  address={Department of Physics and Astronomy, San Francisco State University, 1600 Holloway Avenue, San Francisco, CA 94132, USA}
}

\author{Michelle B. Brochmann}{
  address={Department of Physics and Astronomy, San Francisco State University, 1600 Holloway Avenue, San Francisco, CA 94132, USA}
}

\author{Jay Anderson}{
  address={Space Telescope Science Institute, Baltimore, MD 21218, USA}
}

\author{Melvyn B. Davies}{
  address={Lund Observatory, Box 43, SE-221 00 Lund, Sweden}
}

\begin{abstract}

We summarize results of a search for X-ray-emitting binary stars in
the massive globular cluster $\omega$ Centauri (NGC 5139) using \Chandra\
and \HST.  ACIS-I imaging reveals 180 X-ray sources, of which we
estimate that $45-70$ are associated with the cluster.  We present 40
identifications, most of which we have obtained using ACS/WFC imaging
with \HST\ that covers the central 10\amin \x 10\amin\ of the cluster.
Roughly half of the optical IDs are accreting binary stars, including
9 very faint blue stars that we suggest are cataclysmic variables near
the period limit.  Another quarter comprise a variety of different
systems all likely to contain coronally active stars.  The remaining 9
X-ray-bright stars are an intriguing group that appears redward of the
red giant branch, with several lying along the anomalous
RGB.  Future spectroscopic observations should reveal whether these
stars are in fact related to the anomalous RGB, or whether they
instead represent a large group of ``sub-subgiants'' such as have been
seen in smaller numbers in other globular and open clusters.

\end{abstract}

\maketitle

\section{Introduction}

Globular clusters (GCs) host a variety of binary star systems, formed 
both primordially and dynamically via stellar encounters. These binaries 
play a crucial role in the dynamical evolution of GCs, providing 
an energy reservoir that can delay core collapse for many times the 
half-mass relaxation time \citep[\eg,][]{Fregeau07}. The dense cluster 
environment also dramatically alters the evolution of GC binaries 
\citep[\eg,][]{Ivanova06,Fregeau03,Pooley06,Fregeau08}. X-ray-emitting systems 
have, in particular, emerged as a promising source of information about the 
history of binary formation and destruction in galactic GCs. 

The {\it Chandra X-ray Observatory}'s high spatial resolution and resulting 
sensitivity to point sources makes it possible to obtain nearly complete 
samples of compact accreting binaries in nearby globular clusters.  
The ability to pinpoint sources to $< 1$\asec\ also means that the stars 
responsible for the X-ray emission may be recovered at other wavelengths even 
in the crowded fields of GCs. While the high luminosity X-ray binaries 
($L_x = 10^{36-38}$ erg\,s$^{-1}$) are understood to be accreting neutron 
stars \citep{Brown98,Heinke03a,Heinke03b}, the low X-ray-luminosity sources are 
now known to comprise several distinct populations: cataclysmic variables (CVs), 
quiescent neutron stars (qNS, or qLMXB), millisecond pulsars (MSPs), and 
binaries with chromospherically active stars, \ie\ active binaries
\citep[ABs;][]{Pooley02b,Heinke05,Lugger07,Haggard09}. Of these, only the quiescent 
NSs, with their distinctive soft X-ray spectra, can be identified uniquely on the 
basis of X-ray observations alone \citep{Pooley02a,Rutledge02}. For others, optical 
(or radio, in the case of MSPs) follow-up is essential. 

\wcen\ is the most massive GC in the Milky Way \citep[$4 \times
10^6$\msun;][] {Meylan02}. At 4.9 kpc, it is relatively nearby, making
it possible to detect low-luminosity X-ray sources in modest exposure
times with \Chandra. Its unusually complex stellar populations have
prompted debate as to whether \wcen\ is a GC at all --- it is instead
likely to be the remnant of a dwarf galaxy accreted by the Milky Way
\citep{Bedin04, Gratton04, Piotto05, Villanova07} --- and controversy
continues over the existence of an intermediate black hole in its
core, \eg\ see \citet{Noyola08} vs. \citet{Anderson10}.  Regardless of
\wcen's origins, the binary stars that it contains play a crucial role
in its dynamical evolution and can in turn shed light on the impact
that a cluster has on its binary population.  Here we summarize
results of our search for X-ray-emitting binary stars
in \wcen\ using {\it Chandra X-ray Observatory} and {\it Hubble Space 
Telescope} (\HST). The \Chandra\ results have been reported by \citet{Haggard09} 
and the complete \HST\ results will appear in \citet{Cool10}.

\section{Chandra Observations and Results}

The \Chandra\ observations were made using the Advanced CCD Imaging
Spectrometer (ACIS), whose field of view (FOV) is \about 17\amin \x\
17\amin.  For comparison, the half-mass radius of \wcen\ is 4.2\amin\ 
\citep{Harris96} and its core radius is \rc\ $=$ 2.6\amin\ \citep{Trager95}. 
With a total exposure of \about 70 ksec, we detected 180 sources to a limiting 
X-ray flux of \about 4.3 \x\ $10^{-16}$ erg\,cm$^{-2}$\,s$^{-1}$.  At the 
distance of \wcen\ (4.9 kpc), this corresponds to $L_x$ \about 1.2 \x\
$10^{30}$ erg\,sec$^{-1}$.  Because \wcen\ is very large on the sky,
X-ray sources anywhere in the ACIS-I field can potentially be cluster
members.  However, given the large FOV and faint limiting
flux of the observations, significant numbers of AGN will be present.
After a statistical accounting of AGN as well as foreground stars, we
estimated that $45-70$ of the \Chandra\ sources are associated with the
cluster.  Based on nine optical IDs we projected that perhaps $20-35$ of
the sources were cataclysmic variables (CVs), with most of the
remainder being binaries containing coronally-active stars
\citep[see][for details]{Haggard09}.

Figure \ref{xray_cmd} shows an X-ray color-magnitude diagram for all the \Chandra\
sources for which counts were recorded in three bands: "soft" ($0.5-1.5$ keV),
"medium" ($0.5-4.5$ keV), and "hard" ($1.5-6.0$ keV). Black symbols indicate the
radial offset of each source from the cluster center; colored symbols mark X-ray 
sources for which optical identifications have been obtained. In Fig. \ref{xray_cmd} 
and the descriptions below, we exclude sources for which the identifications suggest 
they are not associated with the cluster (several AGN and a few foreground stars). We 
also exclude objects whose optical signatures are ambiguous.  The complete set of 
optical IDs will be presented by \citet{Cool10}.

\section{HST Observations and Analysis}

The \HST\ data consist of 9 pointings with the ACS Wide Field Camera
(WFC) covering \about 10\amin\ \x 10\amin, approximately centered on
the cluster; this field encompasses 109 of the \Chandra\ sources.  At
each pointing we obtained four F625W (R$_{625}$), four F435W
(B$_{435}$), and four F658N (\ha) exposures.  The broad-band
exposures include one short exposure to measure the bright stars.  To
map the positions of \Chandra\ sources onto the ACS/WFC images, we
corrected the WFC images for distortion and then constructed mosaics
in each filter.  We used the star lists of \citet{Kaluzny96} and
\citet{vanLeeuwen00} to map R.A.\ and Dec.\ onto the mosaic images and
then back onto the original "flt"-format images.

We did the photometric analysis using DAOPHOT/ALLSTAR
\citep{Stetson87} and ALLFRAME \citep{Stetson94} on the "flt"-format
images.  We first extracted \about 20\asec \x\ 20\asec\
"patches" centered on each X-ray source from each of the ACS/WFC
exposures available at the corresponding position.  For most sources,
this meant analyzing a total of 12 images.  Because of the possibility
that an interesting optical counterpart could be missed in a fully
automated process, we carefully scrutinized the error circle region
and iterated several times to insure that all objects identified
within it were real and nothing was missed \citep[details in][]{Cool10}.  
Initially we adopted 1\secspt 0-radius error circles.  Once
several identifications had been made we performed a boresight
correction which enabled us to reduce the radius to 0\secspt 6.
Finally, we constructed \br\ vs.\ \r\ and \hr\ vs.\ \r\
color-magnitude diagrams (CMDs) for each patch.

Once we had constructed CMDs for the patch around each X-ray source,
we carried out a systematic evaluation of potentially interesting
objects.  All objects that did not lie on or very near the main
sequence or giant branch in both the \br\ and \hr\ CMDs were
considered potentially interesting and evaluated for reliability of
the photometry.  We examined them in all the individual images to
check the potential impact of near neighbors, cosmic rays, and
diffraction spikes, and checked how cleanly DAOPHOT removed them from
each of the images.  We also took account of the consistency of
multiple independent measurements in each filter.  We then
assigned a numerical index to represent the quality of the photometry
(0 = best, 3 = worst); here we report only quality 0 and 1 candidates.

Figure \ref{opt_cmd} shows a combined \br\ vs.\ \r\ and \hr\ vs.\ \r\
CMD for 35 of the optical IDs obtained using the \HST\ data.  Black
dots mark the stars in the error circles of these 35 objects.  To
better delineate the turnoff and giant branch we also plot stars
brighter than \r\ = 19 from the full 20\asec \x\ 20\asec\ patch
associated with the qLMXB.

\section{Optical Counterparts}

\begin{figure}
  \includegraphics[height=.4\textheight]{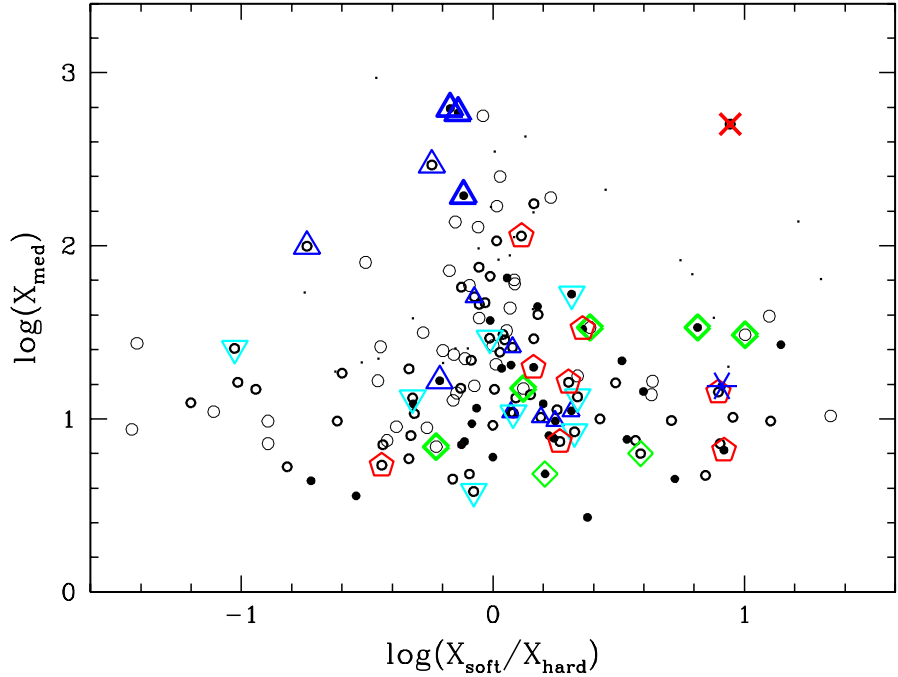}
  \caption{X-ray "color-magnitude" diagram for 164 \Chandra\ sources in \wcen\ with 
  non-zero counts in both hard and soft bands. Round symbols indicate where in the 
  cluster a source lies (large solid dots $=$ core; large open circles $= 1-2$ \rc; 
  small open circles $= 2-3$ \rc; dots $=$ outside 3 \rc). Special symbols indicate 
  optical IDs: a quiescent neutron star (red cross); previously known and newly 
  identified CVs (large blue triangles); less certain new 
  CVs (smaller blue triangles); very faint CV candidates with no \ha\ detection
  (cyan inverted triangles); active binaries, \eg\ BY Dra systems (green  
  diamonds); possible sub-subgiants, also sometimes called red stragglers (red 
  pentagons); and a candidate blue straggler (blue asterisk).  Bold symbols indicate 
  optical counterparts identified by \citet{Carson00} and \citet{Haggard09}, including 
  5 ABs from the \citet{Kaluzny04} variable star catalog.}
  \label{xray_cmd}
\end{figure}

\begin{figure}
  \includegraphics[height=.56\textheight]{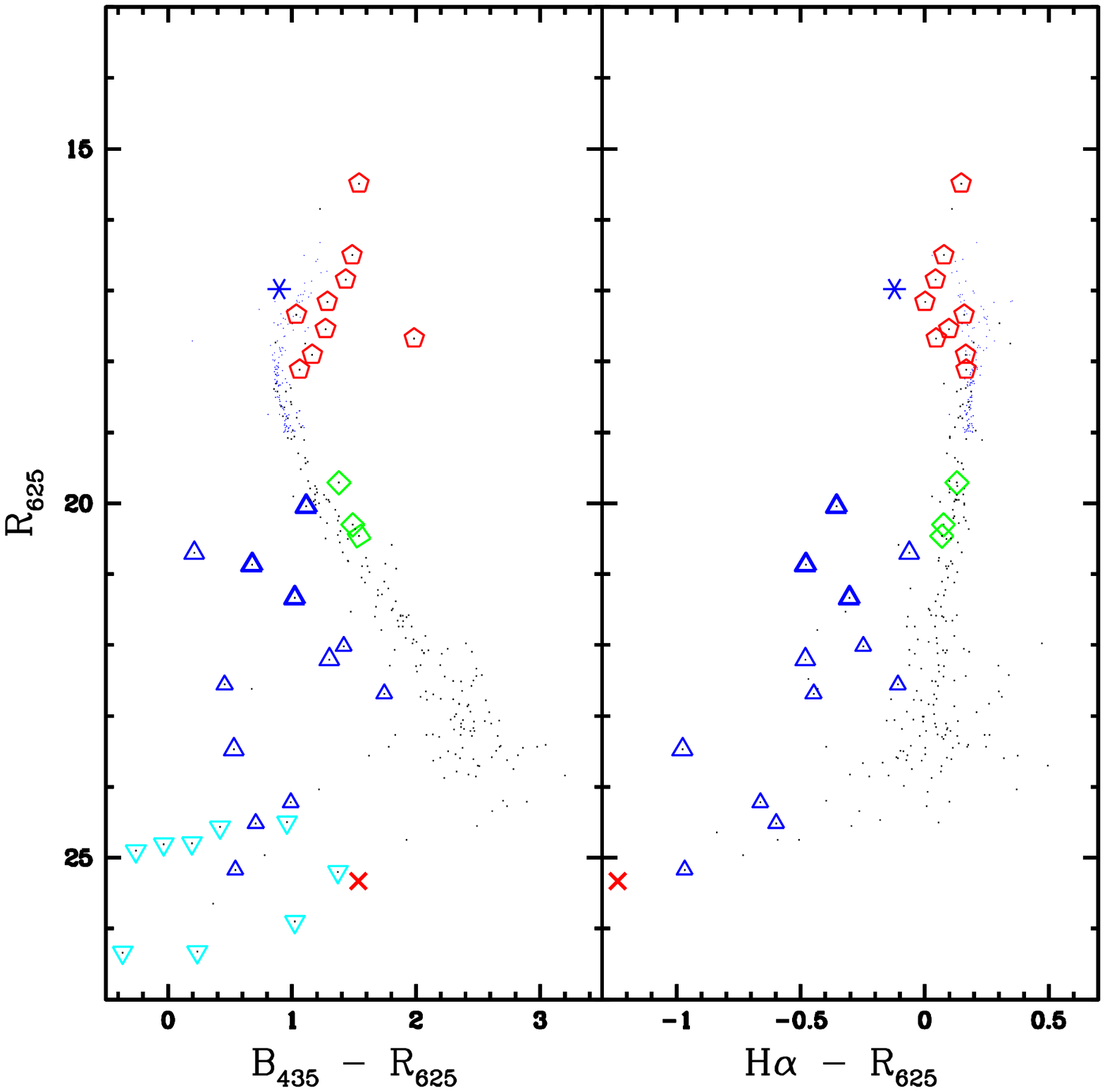}
  \caption{Color-magnitude diagram in B$_{435}$ $-$ R$_{625}$ vs. R$_{625}$ (left 
  panel) and \ha $-$ R$_{625}$ vs. R$_{625}$ (right panel) showing all stars 
  that appear in the error circles of 35 X-ray sources for which promising optical
  identifications have been made.  Black dots represent all the stars in the error 
  circles.  To better delineate the turnoff region and red giant branch, all stars with 
  R$_{625}$ $<$ 19 in the full 20\asec \x\ 20\asec\ patch around the qLMXB are shown as
  blue dots.  Suggested optical counterparts of X-ray sources are marked with 
  symbols as in Fig. \ref{xray_cmd}.  Note that the 5 ABs obtained by matching X-ray
  sources to the \citet{Kaluzny04} variable star catalog are not shown as they either
  lie outside the \HST\ FOV or were not distinctive in the \HST\ data.}
  \label{opt_cmd}
\end{figure}

\subsection{Accreting Binary Stars: Cataclysmic Variables and a qLMXB} 

Thirteen of the X-ray sources have an optical counterpart that is both
blue and \ha-bright (Fig. \ref{opt_cmd}). This combination of signatures is
strongly indicative of a compact accreting binary, with the \ha\
excess attributable to an emission line from an accretion disk and the
blueness to the disk and/or the white dwarf.  One of these is the
object reported by \citet{Haggard04}, which was first identified as a
qNS on the basis of its X-ray spectrum \citep{Rutledge02}; it is marked with a red 
cross in Figures \ref{xray_cmd} and \ref{opt_cmd}.  This is the only X-ray source in 
\wcen\ with both the soft spectrum and the luminosity characteristic of a quiescent 
neutron star (see red cross in Fig. \ref{xray_cmd}).

The remaining 12 optical counterparts that are both blue and
\ha-bright have X-ray colors and luminosities typical of cataclysmic
variables (CVs; see blue triangles).  Three of these (bold blue
triangles) were known from previous studies \citep{Carson00, Haggard04}; 
nine are new.  The best of the new candidates (large blue triangles) are 
all easily confirmed visually as being blue and all individual \hr\ measurements 
are consistent in showing that they are \ha-bright.  The remaining six CV candidates
(small blue triangles) are visually confirmed as either blue or
\ha-bright but have somewhat lower confidence associated with either their \ha\
excess or blueness.  Nevertheless, we think it probable that most, if
not all, of these stars are the optical counterparts of the X-ray
sources and are CVs.

In addition to these 12 CV candidates, we find very faint blue stars in
the error circles of nine of the \Chandra\ sources (see inverted cyan
triangles).  These nine stars are exceedingly faint (\r $= 24.5-26.3$).
Most are seen in \b\ only because they are so blue; main-sequence
stars of comparable \r\ magnitude are below the detection limit.  All
but one were confirmed visually as being blue by blinking \r\ vs.\
\b\ images.  However, as none of these stars are detected in \ha,
they do not appear in the \hr\ diagram (see Fig. \ref{opt_cmd}).
Given that these stars lie in the region of the CMDs generally occupied by 
white dwarfs (WDs), we have considered the possibility that they could be WDs 
that have landed by chance in the \Chandra\ error circles and are unrelated to 
the X-ray sources.  Our statistical analysis \citep{Cool10} suggests
that perhaps one could be such a chance coincidence, but that it is
unlikely that many more could be explained in this way.  We therefore
suggest that these objects are also CVs.  It is important to note that
the lack of \ha\ detections does not necessarily imply that they are
not \ha-bright; they may simply be too faint to be detected, even in
the presence of an \ha\ emission line.

\subsection{Binaries Containing Stars with Active Coronae}

X-ray imaging with \Chandra\ of nearby GCs is also
sensitive enough to pick up binaries containing coronally active
stars.  Falling broadly into this category are the five optical
identifications by \citet{Haggard09} based on comparison of the
variable star catalog of \citet{Kaluzny04} with the \Chandra\ source
list.  These five sources, which include two eclipsing Algols and a
long-term variable, are shown as bold green diamonds in Figure \ref{xray_cmd}.
They do not appear in Figure \ref{opt_cmd}, as four are outside the \HST\ FOV,
and the fifth was not picked up as being distinctive in the \HST\
CMDs.

Narrow-band imaging with \ha\ also enables us to search for the
elevated levels of coronal activity associated with certain types of
binary stars (e.g., BY~Dra and RS~CVn).  In the field, this activity
is typically the result of fast spin rates induced by tidal
synchronization \citep[see][and references therein]{Makarov09}.
However, \ha\ equivalent widths for such stars are much lower than for
CVs, \about $1-3$ Angstroms \citep{Young89}.  Given the \about 80
Angstrom width of the ACS/WFC \ha\ filter, such stars will appear only
very slightly \ha-bright (at best) in the present study.  To limit the
number of false positives in our search, we required that a
star show both an \ha\ signature and also be above the main sequence
in the individual \br\ CMD associated with its patch.  The latter
requirement excludes binaries whose mass ratio is much less than
unity.  The three candidate BY~Dra stars we found in this way are
shown in Fig. \ref{opt_cmd} (green diamonds).  In Fig. \ref{xray_cmd}
it can be seen that, on average, these are relatively faint and
moderately soft X-ray sources.

\subsection{A Blue Straggler or Turnoff Binary?}

One of the candidates (blue asterisk in Fig. \ref{opt_cmd}) appears
above the turnoff, to the blue side of the subgiant branch, a location
which is suggestive of a blue straggler.  The star has a
0.25-magnitude \ha\ excess, which strongly suggests that it is the
X-ray source.  However, the star also lies \simlt 0.75 magnitudes
above the turnoff and thus may instead be a detached binary containing
two turnoff stars.  We suggest that this star probably falls in the
broad category of active stars, in the sense that its X-ray emission
is most likely to be associated with an active corona.  However, we
give it its own symbol in the CMDs to distinguish it as a special case.

\subsection{Sub-subgiants or Anomalous RGB Stars?}

Nine of the candidates we have identified appear to the red side of
the main-sequence turnoff and subgiant and giant branches
(Fig. \ref{opt_cmd}; red pentagons).  Several show signs of \ha\ in
emission, which is strongly associated with enhanced X-ray emission
and supports an association between these stars and the X-ray sources.
Given their location in the CMD, we tentatively identify these stars
as sub-subgiants \citep[SSG;][]{Mathieu03}, also sometimes called red
stragglers.  However, a close inspection of the CMD shows that 7 of
the stars lie along the metal-rich ``anomalous'' red giant
branch \citep{Pancino00,Villanova07}.  Thus it is possible that these
stars are instead a subset of the anomalous RGB stars which for some
reason are unusually X-ray bright.

\section{Discussion}

Using a combination of \Chandra\ and \HST\ imaging in blue,
red and \ha\ filters, we have identified a total of 40 X-ray-emitting
binary stars in \wcen.  Five were found as X-ray counterparts of
variable stars reported by \citet{Kaluzny04} and four had been found
in earlier \Chandra\ and/or \HST\ studies.  The remaining 31 are newly
reported here.

Accreting binary stars make up just over half of the identifications:
one qLMXB and 21 candidate CVs.  The remaining identifications are
evenly split into two broad classes: 9 active binaries and 9 objects
that are possible sub-subgiants.  The active binaries include an
assortment of different types of systems, all of whose X-ray emission
is likely due to active coronae.  These include two eclipsing Algol
systems, three possible BY Dra stars, and a blue straggler.  

Of particular interest among the CV candidates are the nine very faint
blue stars shown as inverted triangles in Figures 1 and 2.  Given the
distance modulus to \wcen, their absolute magnitudes are in the range
$M_{625} = 10.9-12.7$.  This is comparable to the absolute magnitudes of
the short-period CVs recently uncovered in the Sloan Digital Sky
Survey \citep[SDSS;][]{Gansicke09}.  Thus the systems in \wcen\ could
be short-period systems with very low-mass secondaries, as is expected
for very old CVs.  Their positions in the X-ray CMD (see Fig. \ref{xray_cmd})
generally support the CV interpretation.  Alternatively, some of these
objects could be helium white dwarfs with MSP companions; a few such
systems are known in globular clusters \citep[\eg\,][]{Edmonds01}.  
Deeper \ha\ imaging and/or multiwavelength broad-band imaging is needed to 
distinguish between these possibilities.

Given that the faintest CVs detected in this study are at the
detection limit in the both the X-ray and optical images, it is likely
that more CVs remain to be discovered in \wcen.  How many more depends
on the relative numbers of faint vs.\ bright CVs --- a ratio that
depends both on the evolution of CVs and their formation history in
the cluster.  The present study shows that even very faint CVs can be
found in the crowded environs of \wcen; deeper observations should
allow a more complete census to be made.

The present census of active binaries in \wcen\ is undoubtedly very
incomplete.  At X-ray wavelengths we are likely seeing just the tip of
the iceberg.  In the optical, we are hampered by the weakness of the
\ha\ emission lines.  Observations with a narrower \ha\ filter
can help, as demonstrated by the large number of BY~Dra stars
identified by \citet{Taylor01} in NGC~6397 using a \about 20
Angstrom-wide filter.  

Perhaps the most intriguing set of X-ray-emitting stars identified in
this study are the nine that lie redward of the turnoff and giant
branch.  While their close proximity to the evolutionary sequences in
\wcen\ suggests that they are associated with the cluster, proper motions
are needed to be sure.  If spectroscopic observations reveal that some
or all are members of the anomalous RGB, then it will be important to
understand why this subpopulation in \wcen\ is prone to producing
X-ray-bright stars.  If instead these stars turn out to be bonafide
SSGs, then \wcen\ should provide a valuable testing ground for studying
this as-yet poorly understood class of X-ray-emitting binary systems.

\bibliographystyle{aipproc}   

\end{document}